\documentclass[prl,twocolumn,reprint,showpacs,floatfix,superscriptaddress,showpacs]{revtex4-1}
\usepackage{graphicx}
\usepackage{times}
\usepackage{amsmath,amssymb,bm}
\usepackage{dcolumn}
\usepackage{mathtools}
\usepackage{color}
\newcommand{\ii}{\mathrm{i}}
\newcommand{\dd}{\mathrm{d}}

\begin{document}

\title{Low-energy Spin Dynamics of the Honeycomb Spin Liquid Beyond the Kitaev Limit}
\author{Xue-Yang Song}
\affiliation{International Center for Quantum Materials, School of Physics, Peking University, Beijing, 100871, China}
\author{Yi-Zhuang You}
\affiliation{Department of Physics, University of California, Santa Barbara, California 93106, USA}
\author{Leon Balents}
\affiliation{Kavli Institute of Theoretical Physics, University of California, Santa Barbara, Santa Barbara, CA 93106}
\date{\today}
\begin{abstract}
We investigate the generic features of the low energy dynamical spin structure factor of the Kitaev honeycomb quantum spin liquid perturbed away from its exact soluble limit by generic symmetry-allowed exchange couplings.  We find that the spin gap persists in the Kitaev-Heisenberg model, but generally vanishes provided more generic symmetry-allowed interactions exist. We formulate the generic expansion of the spin operator in terms of fractionalized Majorana fermion operators according to the symmetry enriched topological order of the Kitaev spin liquid, described by its projective symmetry group. The dynamical spin structure factor displays power-law scaling bounded by Dirac cones in the vicinity of the $\Gamma$, $K$ and $K'$ points of the Brillouin zone, rather than the spin gap found for the exactly soluble point. 
\end{abstract}
\maketitle

Quantum spin liquids (QSLs) have attracted wide attention due to their intriguing highly entangled nature and exotic properties \cite{savary2016quantum,balents2010spin}.  Amongst the simplest and most interesting QSLs are those with intrinsic topological order, which are of particular interest as potential platforms for quantum computing possessing intrinsic protection from decoherence\cite{kitaev2003fault}.   A prominent feature of topological phases and QSLs in general is the fractionalization of electrons or spins into other particles, Majorana fermions for example.   An unequivocal observation of this fractionalization is a key goal of the QSL field.

Recent theory and experiment have unveiled the exciting prospect of achieving this objective in highly anisotropic spin-1/2 magnets on honeycomb lattices, including Na$_2$IrO$_3$, Li$_2$IrO$_3$, and $\alpha$-RuCl$_3$\cite{jiang2011theory, reuther2011theory,knolle2014theory,kim2015theory, alpichshev2015experiment, sandilands2014experiment, plumb2014experiment,experiment2,experiment1,experiment4,singh2010antiferromagnetic,choi2012spin,singh2012}.   Two key theoretical works presaged this experimental venue.  First, a seminal paper by Kitaev\cite{kitaev2006anyons} introduced a simple near-neighbor spin Hamiltonian on this lattice, possessing a gapless $\mathbb{Z}_2$ QSL phase, the excitations of which are massless relativistic (i.e. linearly dispersing) Majorana fermions and gapped bosonic ``fluxes''\footnote{Kitaev also demonstrated other phases in the limits of strong anisotropy and upon application of a particular orientation of magnetic field.  This is, however, not relevant to our discussion.}.  Second, Jackeli and Khaliullin\cite{jackeli2009mott} showed that Kitaev's anisotropic interactions arises naturally from certain superexchange processes in strongly spin-orbit coupled transition metal compounds.   These two developments spurred the search for Kitaev's QSL in this context in the laboratory.

\begin{figure}[htbp]
 \centering
\includegraphics[width=0.38\textwidth]{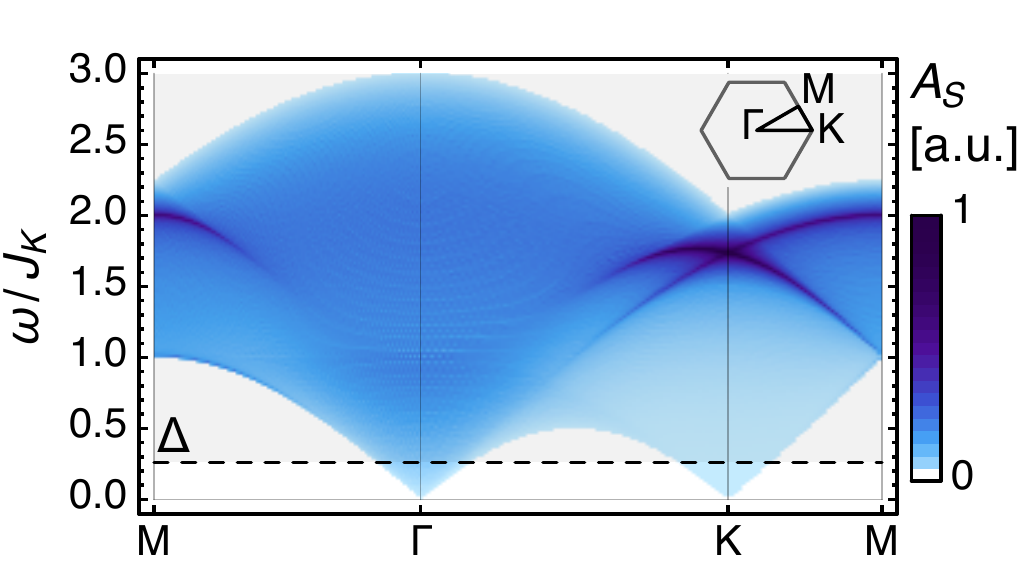}
 \caption{The spectral function along high symmetry line at the isotropic point. The simplest form as listed in eq~\eqref{eqn:simple} is used for calculation performed on a honeycomb lattice with $240\times 240$ unit cells. The dashed line marks the gap $\Delta=0.262J_K$ of flux excitations, above which the leading contribution from the unperturbed Kitaev spin liquid will dominate. }
 \label{fig:spectral}
 \end{figure}

\begin{figure}[htbp]
\centering
\includegraphics[width=0.46\textwidth]{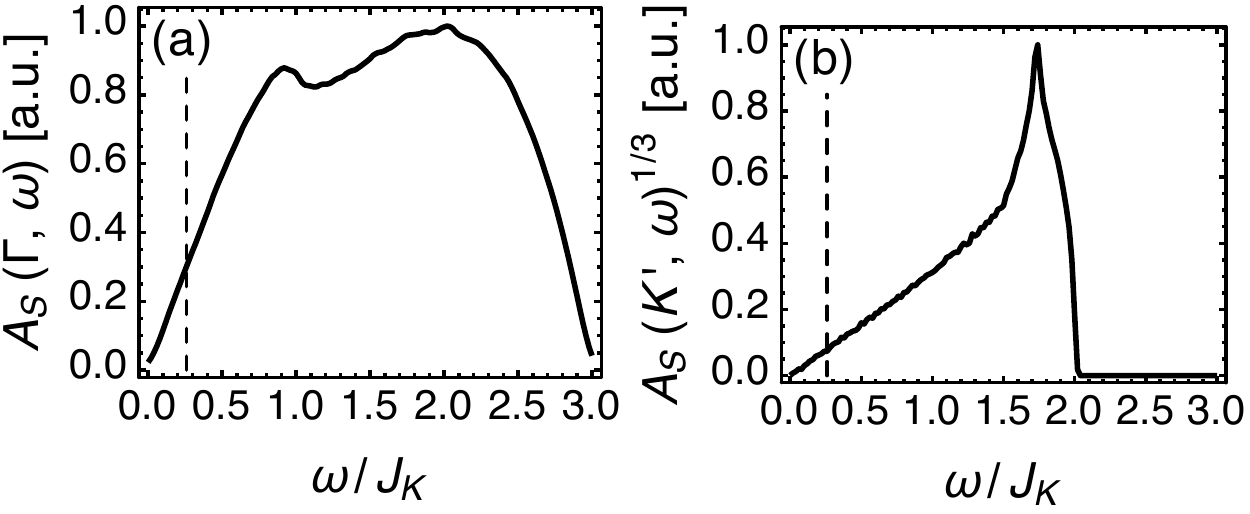}
\caption{The spectral function versus frequency at (a) the $\Gamma$ point $\bm q=0$ and (b) the $K'$ point $\bm q\sim 2\bm q_0$. The cube root of $A_S(K',\omega)$ is plotted in (b) to demonstrate the scaling behavior $A_S(K',\omega)\sim\omega^3$. The dashed line marks the flux gap $\Delta=0.262J_K$.}
\label{fig:spectral_curve}
\end{figure}

In this paper, we address a key experimental signature of any magnet, the dynamical spin structure factor, for the Kitaev QSL. The dynamic spin response can be measured using conventional experiment techniques such as inelastic neutron scattering and electron spin resonance.  It is given, at zero temperature, by
\begin{equation}
  \label{eqn:sf}
S^{\mu\nu}_{ij}(t)=\langle 0| T(\sigma^{\mu}_{i}(t)\sigma ^{\nu}_{j}(0))|0\rangle,
\end{equation}
where $\sigma^\mu_i$ is the Pauli operator representing the $\mu^{\rm th}$ component of the spin at site $i$ of the lattice, and the arguments indicate the usual Heisenberg time evolution.
Previous studies demonstrated that for Kitaev's exactly soluble model $S^{\mu\nu}_{ij}$ vanishes between all but the neighbor pair of spins \cite{Shankar2007}.   Moreover, the dynamical spin response exhibits a spin gap -- a non-zero interval of frequency around zero in which the spectral weight vanishes --  despite the existence of gapless excitations \cite{knolle2014dynamics, knolle2015SF}. These remarkable properties arise due to the exact integrability of the Kitaev model.  Here we ask the important question whether the apparent spin gap persists when moving away from the exactly solvable point in the critical spin liquid phase\cite{tikhonov2011sf,mandal2011sf}.  We find that the existence of the gap as a robust property of the QSL phase relies critically upon internal symmetries: it is present in the Heisenberg-Kitaev model but {\em not} in the generic model allowed by physical symmetries in actual materials.  In the latter case we obtain {\em universal} power-law spectral weight at low energies, as shown in Figs.~\ref{fig:spectral},~\ref{fig:spectral_curve}.  This, rather than the gap found in Ref.\onlinecite{knolle2014dynamics}, is the expected behavior should the Kitaev QSL be realized in actual experiments.

To obtain these results, we follow standard arguments of low energy effective field theory.  The low energy field theory  in the gapless phase of Kitaev model is a single cone of massless Dirac fermions (a convenient formulation for two Majorana cones).  Physical operators may be expanded in the primary field and descendents of the field theory, here the Dirac fields, and we expect all terms consistent with symmetry to generically appear in this expansion.  Due to the fine-tuned nature of the exactly soluble point, many coefficients in this expansion vanish there, but will become non-zero if symmetry allows. 
In the case of a QSL phase, the analysis of permitted terms is subtler because the physical symmetries are intertwined with emergent gauge transformations of the non-local fermions.  This is described by the mathematical structure of projected symmetry groups (PSGs)\cite{senthil2000psg,balents2005psg, senthil2005psg,lawler2008psg,you2012doping}.  Here, we use the PSG analysis of Ref.\onlinecite{you2012doping} to find the low energy contributions to the microscopic spin operator, and from this deduce the dynamical spin correlations.  To complement the PSG analysis, we also study the problem directly by perturbation theory away from the soluble Kitaev point, thereby obtaining the scaling of the prefactors $f_j$'s in Eq.~\eqref{eqn:general} of the symmetry-allowed terms.

\begin{figure}
\includegraphics[width=0.48\textwidth]{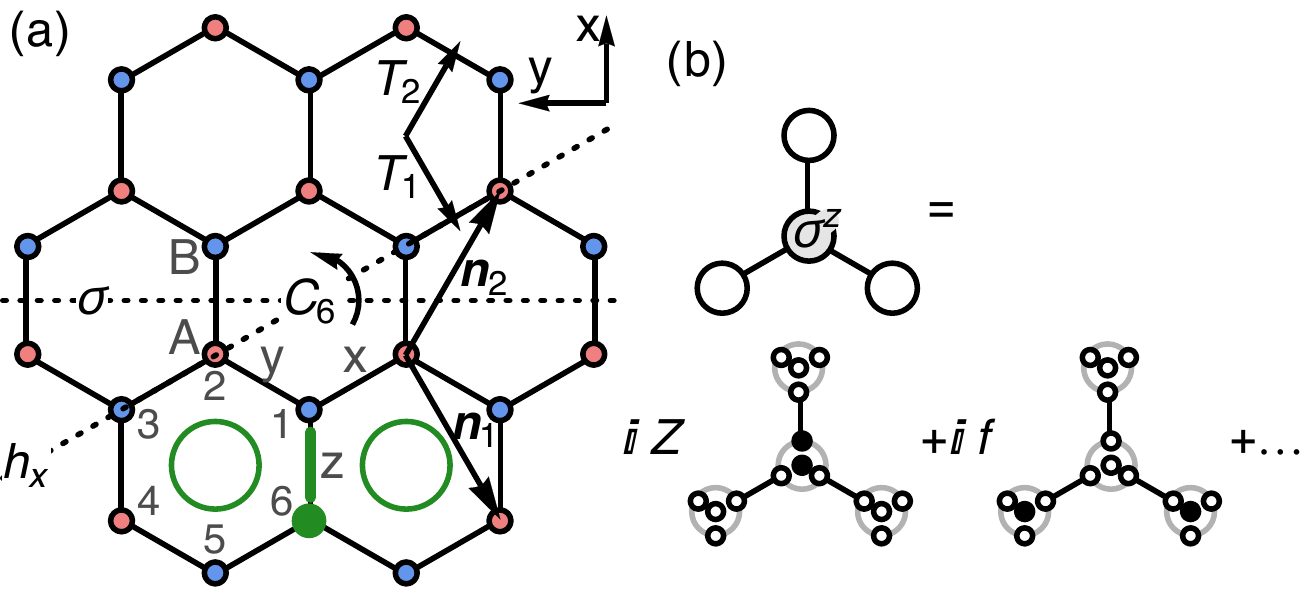}
\begin{center}
\caption{(a) The honeycomb lattice for Kitaev model. A, B denote sublattice index. A spin $\sigma_6^z$ (in green) on the ground state changes the $1 \text{-}6$ bond operator eigenvalue, creating fluxes in two plaquettes sharing that bond. The symmetries $T_1,T_2,C_6,\sigma$ and $h_x$ are displayed. 
$(x,y)$ coordinate system is shown on the upper right conner. 
(b) Spin operators schematically expanded in terms of Majorana fermion operators. Each site contains four Majorana modes (small dots), where the central one is $c_i$ and the surrounding three are $c_i^\mu$'s. Each black dot represents a corresponding Majorana operator present. 
} 
\label{fig:Honeycomb}
\end{center}
\end{figure}

We begin by recapitulating Kitaev's model and its solution.  It consists of interacting spin-1/2 moments on a honeycomb lattice. 
Original hamiltonian reads 
\begin{equation}
 H_0= J_K \sum_\mu\sum_{\langle ij \rangle^\mu} \sigma^\mu_i\sigma^\mu_j, \label{eq:4}
\end{equation}
where $\langle ij\rangle^\mu$ denotes the neighbor sites $i,j$ whose bond direction is labeled by $\mu=x,y,z$ (Fig.~\ref{fig:Honeycomb}(a)). There is a local conserved quantity for each plaquette $W_p=\sigma^x_1\sigma^z_2\sigma^y_3\sigma^x_4\sigma^z_5\sigma^y_6$ (Fig.~\ref{fig:Honeycomb}(a)), the set of which on all plaquettes is a set of good quantum numbers labeling energy eigensectors. We say a $\pi$-flux is present on a plaquette $p$ if $W_p=-1$. The solution to Kitaev model is well known through the mapping to a free fermion  Hamiltonian 
\begin{equation}
\label{eqn:majorana_Hamiltonian}
H_0=J_K \sum_{\mu}\sum_{\langle ij\rangle^\mu}\ii c_i {\hat u}_{\langle ij\rangle^\mu}c_j,
\end{equation}
if we write each spin as the product of Majorana operators $\sigma^a_i = \ii c_ic_i^\mu$ (physical subspace satisfies $c_ic_i^xc_i^yc_i^z=1$) and define bond operator $\hat u_{\langle ij\rangle^\mu}=\ii c^\mu_ic^\mu_j$. The bond operators with eigenvalues $\pm1$ commute with the Hamiltonian and the product of them around a plaquette $p$ is $W_p$. Identifying the eigenvalue of $\hat u_{\langle ij\rangle^\mu}$ as $\mathbb Z_2$ gauge field, one can obtain the ground state that exists in the zero-flux subspace by setting $\hat u_{\langle ij\rangle^\mu}$'s to 1 and diagonalizing the Hamiltonian. Complementary to the static flux sector, the Majorana fermions $c_i$'s living on-site span the matter fermion (spinon) sector.

From this structure, the short-range and gapped nature of the spin correlations for the soluble model follows directly.  When applying a spin operator to the ground state, a spinon excitation is created with two fluxes (illustrated in Fig.~\ref{fig:Honeycomb}(a) in green).  For $\mathbb{Z}_2$ topological order, there are two bosonic anyons, {\sf e} and {\sf m}, corresponding to fluxes, and one fermionic anyon, $\varepsilon$, which is the $c_i$ fermion.   In this language, $\sigma \sim {\sf e} {\sf m} \varepsilon$, which is consistent with general rules since $\varepsilon$ carries both ``electric'' and ``magnetic'' gauge charge making the combination of three anyons gauge invariant. Thus a spin gap appears since there is a finite energy difference between the ground state and the lowest eigenstate of the two-vison sector. Infinitely massive visons render the spin correlators short-ranged.   Yet both properties may be changed if contributions to the spin of the form $\sigma\!\sim\!\varepsilon\varepsilon$ arise, which appears natural based on the gauge structure alone.  To see how they arise, we first consider the structure of states and operators under perturbations, then turn to a general symmetry based analysis.  

{\em Unitary transformation analysis:} Perturbation theory defines a unitary mapping $U=e^{\ii S}$ from eigenstates of the pure Kitaev model $H_0$ to those of the perturbed one $H=H_0+V$, for example the exact ground state $|\psi\rangle = U |\psi_0\rangle$, where $|\psi_0\rangle$ is the unperturbed ground state.  One can find $U$ order by order by demanding that the rotated Hamiltonian $\tilde{H} = U H U^{-1}$ has vanishing off-diagonal matrix elements between low-energy eigenstates of $H_0$, e.g. to first order $\ii S=\sum_{n\neq m} \frac {P_n V P_m}{E_n-E_m}$, where $P_n$ is the projection operator onto the $n$th energy eigenspace.  Formally the higher order terms can be found using Baker-Hausdorff formula: they involve more powers of $V$ separated by projection operators and the corresponding energy differences in the denominators.  Because the energies of eigenstates of $H_0$ with non-uniform fluxes are non-trivial, we are not able to evaluate this explicitly.  However, we can understand the general structure of the expansion.  Moreover, because the Kitaev QSL is a stable phase, $U$ is well-behaved and $S$ defined in this way is a sum of quasi-local operators, at least when restricted to act (on the right) on low energy states.

In general, we can separate any physical operator, including $U$, into a sum of terms which modify the flux on $2k$ sites: $U = \sum_{k=0}^\infty U_{2k}$, where, since $U$ is physical, only an even number of fluxes can be changed.  
We may understand $U_{2k}$ as the mixing $2k$ virtual fluxes into the interacting ground state.

Spin operators transform accordingly under this procedure:
\begin{equation}
  \sigma_i^\mu\rightarrow \tilde\sigma_i^\mu \equiv U^\dagger\sigma_i^\mu U^{\vphantom\dagger}= \sum_{k,k'} U_{2k'}^\dagger \sigma_i^\mu U_{2k} = \sum_k \tilde\sigma_{i,2k}^\mu .\label{eq:1}
\end{equation}
We observe that since $\sigma_i^\mu$ modifies two fluxes, terms with $|k-k'|=0,1$ induce contributions to $\tilde\sigma_{i,0}^\mu$.  The physical picture behind is that the exact ground state of the perturbed system contains terms with two virtual fluxes, which are annihilated by the spin operator.  The resulting state is no longer orthogonal to all the exact zero flux eigenstates.  Thus the transformed spin operator has an expression of the form (Fig.~\ref{fig:Honeycomb}(b))
\begin{eqnarray}
\label{eqn:general}
\tilde\sigma _i^{\mu}=\underbrace{\ii Zc_ic_i^\mu+\cdots}_{\tilde\sigma^\mu_{i,2k>0}}+\underbrace{ f^\mu_{ijk} \ii c_{j}c_{k}+\cdots}_{ \tilde\sigma^\mu_{i,0}}\; ,
\end{eqnarray}
where $Z=1$ for the ideal Kitaev model but is reduced by perturbation.  All other terms become non-zero with perturbations.  The first set of bracketed terms create fluxes exciting modes above a finite energy threshold. We concern here the latter terms, which consist of matter fermions alone and hence create no fluxes and induce only low-energy excitations.

{\em Generic spin operator form from symmetry and gauge constraints:}   We first consider the constraints on the terms which may arise, focusing on the maximal physical symmetry group generated by the following operations
: translations along two basis directions ($T_1,T_2$), time reversal($\mathcal T$), $C_6$ (a 6-fold rotation plus mirror reflection across honeycomb plane) and $\sigma$ symmetry (mirror reflection across the line orthogonal to the z-bond) (Fig.~\ref{fig:Honeycomb}(a), c.f. Ref.~\onlinecite{you2012doping}). In addition, we adopt $h_x$ symmetry (a $\pi$ rotation around the direction of x-bond, equivalent to $\sigma C_6$).  The transformation of Majorana operators should conform to the symmetry enriched topological order of the Kitaev spin liquid, described by PSGs\cite{you2012doping}. 

With these symmetries, we construct combinations of matter fermions compatible with the transformations of spin operators.  A basic constraint is gauge invariance imposed on any physical operator.   The $\mathbb{Z}_2$ gauge transformation  induced by $\eta_i=\pm 1$ takes $u_{\langle ij\rangle}\rightarrow \eta_iu_{\langle ij\rangle}\eta_j$ and $c_i\rightarrow \eta_ic_i$.    Any product of an even number of matter fermions $c_i$, while not gauge invariant on its own, can be made so by multiplying it by a string operator ${\mathcal U}_{i_1i_2}=\prod_{j: i_1 \rightarrow i_2} \hat u_{\langle j_n,j_{n+1}\rangle}$, connecting sites $i_1,i_2$ of Majorana fermions.  It is possible to uniquely restore strings in the low energy Hilbert space with $W_p=+1$, so in the following we adopt a notation with implicit strings (they can be readily restored when desired).

Having understood gauge invariance, we consider the symmetry transformations of matter fermion products.  Consider first time-reversal, under which matter fermion operators on the $A$ sublattice are invariant but those on the $B$ sublattice change sign\cite{you2012doping}.   Since a spin is odd under $\mathcal T$, a corresponding product should consist of an even/odd number of matter fermions on each sublattice if the total number of fermions satisfies $N \equiv 2/0\!\mod \!{4}$, respectively (note the imaginary unit needed when $N\equiv 2\!\mod \! {4}$ to ensure hermiticity).

Thus the smallest appropriate number of matter fermions is two, which must live on the same sublattice.  Taking into account the cyclic structure of spin components (which permute under rotations) and the antisymmetry of fermion bilinears, we postulate the form
(See Fig.~\ref{fig:Honeycomb}(b))
\begin{eqnarray}
\label{eqn:simple}
\tilde\sigma_{i,0}^\mu \sim \frac{\ii}{2} f \epsilon^{\mu\nu\lambda} c^{\vphantom\dagger}_{i+ s_i \hat{\nu}} c_{i+s_i \hat{\lambda}} .
\end{eqnarray}
Here $\hat{\mu}$ is the vector in the $\mu$ direction from the A to B site, and $s_i = +1 (-1)$ for the A (B) sublattice.  It is straightforward to check(Appendix A \footnote{See supplementary material for the discussion on general low-energy forms of spins and the illustration of the flux change in perturbation analysis for specific kinds of perturbations.})
that this form is consistent with all the symmetries.   Restoring the gauge string connecting sites $i+ s_i \hat{\nu}$ and $i+ s_i \hat{\lambda}$, we obtain a gauge invariant expression which can be rewritten in terms of bare spin operators, to wit
\begin{equation}
\label{eqn:simple_form}
  \tilde\sigma_{i,0}^\mu \sim -f \sigma_i^\mu\prod_{\nu \neq \mu} \sigma_{i+s_i\hat{\nu}}^{\nu}.
\end{equation}
Though this operator involves three spins, it is quadratic in the fermions and is expected to give the largest low energy contribution.  Terms with more fermion operators or from further separated sites that give subdominant contributions, are discussed in the Appendix A.

We proceed to check how the form arises in perturbation theory.  Consider the effect of two perturbations on the Kitaev model\cite {Chaloupka2010, Rau2014,singh2012,jiang2011theory}, namely the Heisenberg interaction  $V_H=J_H\sum_{\langle ij\rangle} {\bm {\sigma}}_i\cdot {\bm{\sigma}}_j$ and  the ``cross" term $V_c=J_c\sum_{\mu(\nu\gamma)}\sum_{\langle ij\rangle ^\mu}(\sigma^\nu_i\sigma^\gamma_j+\sigma^\gamma_i\sigma^\nu_j)$ ($\nu,\gamma$ are the remaining directions).   With either {\em one} of the above terms, one can show accidental symmetries cause the spin gap to remain unbroken (i.e. $f$$=$$0$) --  see Appendix B.   However, when both are present, we find that the contribution in Eq.~\eqref{eqn:simple_form} is induced at {\em fourth} order.  Specifically, the sequence of two $J_H$ and two $J_c$ perturbations induces the product $\sigma_2^y\sigma_3^z\sigma_1^x\sim(\sigma_1^x\sigma _2^x)(\sigma_1^x\sigma_2^z)(\sigma_1^x\sigma_3^x)(\sigma_1^x\sigma_3^y)\sigma_1^x$. 
From this, we estimate that the factor $f$ in Eq.~\eqref{eqn:simple} scales as $\frac{J_H^2J_c^2}{J_K^4}$.  


{\em Low-energy weight of the dynamical spin structure factor:} With the low-energy component of the spin operator identified in Eq.~\eqref{eqn:simple}, we can calculate its contribution to the dynamical spin structure factor $S_{ab}^{\mu\mu}(\bm q, \omega)\equiv\frac{1}{N}\sum_{i,j}\int_{-\infty}^{+\infty}\dd t S_{ia,jb}^{\mu\mu}(t)e^{\ii\omega t -\ii{\bm q}\cdot ({\bm r}_i -{\bm r}_j)}$, where $N$ is the number of unit cells in the lattice. We refine the definition of Eq.~\eqref{eqn:sf} by introducing the sublattice indices $a,b=A,B$ apart from the unit cell indices $i,j$. 
In this section, we will show that the low-energy weight of the spin correlators exhibits power law behavior in frequency. 
The single-particle spectrum of spinons in the zero-flux sector is $E(\bm q)=|s_{\bm q}|$ where $s_{\bm q}=J_K(1+e^{\ii\bm q\cdot \bm n_2}+e^{-\ii\bm q\cdot\bm n_1})$ ($\bm n_1$, $\bm n_2$ are basis vectors in Fig.~\ref{fig:Honeycomb}(a)). 
There are two Majorana cones located at the $K$ and $K'$ points ($\pm\bm q_0$) at the Brillouin zone corners, which can be combined into a single cone of Dirac fermions.
Expanding the matter fermion field around the Dirac point, we can write
\begin{equation}
\label{eq:3}
c_i =
\left\{ \begin{array}{ccc} \psi_A({\bm r}) e^{\ii\bm q_0\cdot {\bm r}} +\rm{h.c.}& \qquad & i \in A, \\  \psi_B({\bm r}) e^{\ii\bm q_0\cdot {\bm r}} +\rm{h.c.}& \qquad & i \in B,\end{array} \right.
\end{equation}
where $\psi_a (\bm r)$ is a slowly varying Dirac field (we take $\bm r$ to lie at the hexagon center, i.e. ${\bm r}_i = {\bm r} + \hat{x}$ for $i \in A$ and ${\bm r}_i = {\bm r} - \hat{y}$ for $i \in B$). 
Inserting this into the Hamiltonian in Eq.~\eqref{eqn:majorana_Hamiltonian} and gradient expanding, we obtain the low-energy dynamics of matter fermions described by the action
\begin{equation}
\label{eqn:action}
\begin{split}
S&=\int \dd\tau \dd^2\bm r\, \psi^\dagger[\partial_\tau-v(\sigma_x\ii\partial _x+\sigma_y\ii\partial_y)]\psi  \\
&=\int \dd\omega_n \dd^2\bm q\, \psi^\dagger_{\bm q,\omega_n}(-\ii\omega_n+v\bm \sigma\cdot  {\bm q})\psi_{\bm q,\omega_n},
\end{split}
\end{equation}
where $\psi=(\psi_A,\psi_B)^\intercal$ and the Fermi velocity $v=\sqrt {3}J_K/2$ (The definition of coordinates is indicated in Fig.~\ref{fig:Honeycomb}(a)). 
The field theory is conformally invariant, and the fermion fields scale with length $L$ as $\left[ \psi(\bm r,\tau) \right] = L^{-1}$.  We can decompose the low energy spin operator in Eq.~\eqref{eqn:simple} similarly using Eq.~\eqref{eq:3}.  With some algebra, we obtain the form
\begin{equation}
\begin{split}
\label{eqn:minimal form}
\sigma^{\mu}_{i\in a}\sim \hat M_a^{\mu}(\bm r)+(\ii \hat N_a^{\mu}(\bm r) e^{-\ii \bm q_0\cdot \bm r}+\text {h.c.}),\\
\hat M_a^{\mu}=\psi^\dagger m_a\psi, \quad\hat N_a^{\mu}=\psi^\intercal \bm n_a^{\mu}\cdot\bm \nabla\psi,
\end{split}
\end{equation}
where 
$m_a$ and $\bm n_a^{\mu}$ are two-by-two diagonal matrices, and we used $2{\bm q}_0 = - {\bm q}_0$ up to a reciprocal lattice vector.  Simple specific forms for these matrices in terms of $f$ and ${\bm q}_0$ are obtained by starting with Eq.~\eqref{eqn:simple}, but we expect them to be renormalized generally by higher order terms.  Both the simple and general symmetry-allowed forms are given in the Appendix A.  Note the presence of the gradient in $\hat{N}^\mu_a$: this cannot be avoided because time-reversal symmetry requires the sublattice degrees of freedom be in a  symmetric state, so that the Pauli exclusion principle for this two-particle creation operator forces the orbital wave function to be odd parity; no such requirement applies to the density operator $\hat M_a^\mu$.  

At this point, the scaling of low energy spin correlations is evident.  Using the dimension of $\psi$, the two-point functions of $\hat{M}_a^\mu$ and $\hat{N}_a^\mu$ scale as $\frac{1}{L^4}$ and $\frac{1}{L^6}$, respectively.  The Fourier transformation in 2+1 dimensions adds 3 powers of $L$$\sim$$\frac{1}{\omega}$, so that $S({\bm q}$$\approx$$0,\omega)$$\sim$$|\omega|$ corresponding to $\hat{M}_a^\mu$ correlations, and  $S({\bm q}$$\approx$$\pm {\bm q}_0,\omega)$$\sim$$|\omega|^3$ corresponding to $\hat{N}_a^\mu$ correlations.  

Beyond scaling one obtains the exact low-energy forms by calculation in reciprocal space ($\mu\mu,ab$ are suppressed):
 \begin{equation}
 \begin{split}
 &S(\bm k,\ii\omega_n)\sim\int\dd\omega_1 d^2\bm k_1 \mathrm{Tr}[m_a G(\bm k+\bm k_1,\ii(\omega_n+\omega_1))\\
 &m_b G(\bm k_1, \ii \omega_1)], \\
 &S(\bm q_0+\bm k,\ii\omega_n)\sim \int \dd\omega_1 \dd^2\bm k_1 \mathrm{Tr}\{[\bm n_b^{\mu}\cdot (2\bm k_1-\bm k)] G(\bm k_1,\ii\omega_1)\\
 &[(\bm n_a^{\mu})^\dagger\cdot (2\bm k_1-\bm k)] G^\intercal(\bm k-\bm k_1,\ii(\omega_n-\omega_1))\},
 \end{split}
 \end{equation}
 where $|\bm k|$$\ll$$|{\bm q}_0|$.  The aforementioned scaling follows immediately using $G(\bm k, \ii\omega_n)$$\equiv$$-\langle \psi \psi^{\dagger}\rangle_{\bm k,\omega_n}$$=$$\frac{1}{\ii\omega_n-v\bm \sigma\cdot {\bm k}}$, which is of dimension $\frac{1}{\omega}$, by rescaling $\bm k_1(\omega_1)$$\rightarrow$$\frac{\bm k_1}{\omega_n}(\frac{\omega_1}{\omega_n})$.  One obtains $S(\bm k,\ii\omega_n)$$\sim$$|\omega_n| \tilde{\mathcal{S}}(\frac{v\bm k}{|\omega_n|})$ and $S(\bm q_0+\bm k,\ii\omega_n)$$\sim$$|\omega_n|^3 \tilde{\mathcal{S}}(\frac{v\bm k}{|\omega_n|})$.

Fig.~\ref{fig:spectral} shows the numerical results of spectral function $A_S(\bm q,\omega)$$=$$-\sum_{\mu}$$\sum_{a,b}2\text {Im}[S^{\mu\mu}_{ab}(\bm q,\omega+\ii 0_+)]$ calculated based on the simple forms for $m_a$ and $\bm n_a^{\mu}$ at the isotropic point.  One observes zero low energy spectral weight outside ``Dirac cones'' centered at $\Gamma$ and $K$, i.e. for $\omega$$<$$v|\bm k|$.  Direct inspection of the frequency dependence in Fig.~\ref{fig:spectral_curve} confirms the expected $\omega$ and $\omega^3$ behaviors at $\Gamma$ and $K$, respectively.   Besides this dominant contribution, there will be additional ones arising from products of more than two matter fermion operators, which give larger powers of frequency since every $\psi$ field adds one $\omega$ factor by dimensional analysis. Away form the isotropic limit, i.e. coupling strengths are unequal on bonds for different directions, the Majorana points $\pm{\bm q}_0$ will be shifted away from the Brillouin zone corners, but the scaling of low energy spin correlations, $S(\bm q$$\approx$$0, \omega)$$\sim$$|\omega|$ and $S(\bm q$$\approx$$\pm 2\bm q_0, \omega)$$\sim$$|\omega|^3$ still hold, since $\mathcal T$ still dictates that two-fermion product contains sites on the same sublattice.
 
 
The spectral function $S(\bm q,\omega)$ is of course measured by inelastic neutron scattering.  It also describes the longitudinal nuclear spin relaxation rate $\frac{1}{T_1}$ in a nuclear magnetic resonance (NMR) experiment, in which $\frac{1}{T_1}$ is proportional to the local spectral density of spin fluctuations, which is the momentum integral of $S(\bm q,\omega)$, with $\omega$ at the (low) NMR frequency.  This work predicts temperature scaling $\frac{1}{T_1}$$\sim$$T^3$, in contrast to the ideal Kitaev spin liquid, where the relaxation rate follows the activated behavior $\sim$$e^{\frac{-\Delta}{T}}$ due to the spin gap $\Delta$.  Finally, our result for the uniform component of the spin operator corresponds, if summed over all sites, to  the mass term introduced by Kitaev\cite{kitaev2006anyons} that upon introducing a magnetic field, converts the gapless spin liquid to a gapped non-abelian one.  For the soluble model, this gap is cubic in the applied field and appears only if the field couples to all three spin components.  Our result implies that, generically, the induced gap is {\em linear} in the field and exists for any field orientation.

{\em Conclusion:} We showed that generically the low-energy spectral weight of spin structure factor in the gapless Kitaev QSL phase on honeycomb lattice is non-vanishing, in contrast to the special case of the soluble point.   The results illustrate a general effective field theory approach, which could be applied to any gapless QSL based on a PSG analysis.  In particular for the Kitaev spin liquid we find spectral weight linear (cubic) in $\omega$ near the origin (zone boundary) in momentum space, filling a Dirac-cone-like structure with a sharp spectral edge.   These predictions, and not those of the exactly soluble model, describe the proper low energy behavior of any Kitaev spin liquids which might be found experimentally.  We note that some previous works\cite{tikhonov2011sf,mandal2011sf}\ have discussed the appearance of power-law correlations upon perturbing the Kitaev model, but our results are distinct and more general.\footnote{Ref.\cite{tikhonov2011sf} considered only effects of a field. Ref.\cite{mandal2011sf} showed the absence of low-energy weight for the Heisenberg perturbation, but otherwise gave an incomplete criterion for lifting the gap and did not obtain the effective low energy spin operator or its correlations.}     It would be interesting to test these predictions numerically, e.g. by DMRG calculations (One can include a term $H'$$= $$\Delta \sum_p W_p$ to raise the flux gap by $\Delta$$>$$0$ and open a larger window to observe the low energy weight).

{\em Acknowledgements.}  X.-Y. Song thanks G. Khaliullin, F. Wang and G. Hal\'asz for helpful discussions. This work was supported by the NSF grant no.~DMR-15-06119 (L.B.), the David and Lucile Packard foundation (Y.Y.), the School of Physics at PKU and the 973 Program grant No.~2014CB920902 (X.-Y.S) and benefitted from the facilities of the KITP, supported by NSF grant PHY11-25915.

\appendix*
\section{Supplementary Material for: Low-energy Spin Dynamics of the Honeycomb Spin Liquid Beyond the Kitaev Limit}
   \section{Appendix A: General forms of low-energy spin operator components}

\begin{table}[htb]
\caption{ The transformation of spin and Majorana operators under symmetry operations (we ignore the site label here. Complete transformation is obtained after restoring the corresponding site transformation).}
\begin{center}
  \begin{tabular}{@{} cccccc @{}}
    \hline
    operator & &$C_6$& $\sigma$ & $\mathcal{T}$ & $h_x$ \\ 
    \hline
    $\sigma _{A/B}^x$ &$\rightarrow$& $\sigma _{B/A}^z$ & $-\sigma _{B/A}^y$ & $-\sigma _{A/B}^x$ & $-\sigma _{A/B}^x$ \\ 
    $\sigma _{A/B}^y$&$\rightarrow$&  $\sigma _{B/A}^x$& $-\sigma _{B/A}^x$& $-\sigma _{A/B}^y$ & $-\sigma _{A/B}^z$\\ 
    $\sigma _{A/B}^z$ & $\rightarrow$&$\sigma _{B/A}^y$ & $-\sigma _{B/A}^z$ & $-\sigma _{A/B}^z$ & $-\sigma _{A/B}^y$ \\ 
    \hline
    $c_A$ &$\rightarrow$&  $c_B$ & $c_B$ & $c_A$ & $-c_A$ \\ 
    $c_B$ &$\rightarrow$&  $-c_A$ & $-c_A$ & $-c_B$ & $-c_B$ \\ 
    \hline
  \end{tabular}
\end{center}
\label{tab:transform}
\end{table}

  In this appendix we first examine the constraints that the spatial symmetries impose on the matter fermion operator products to represent a certain spin at the isotropic point. The transformation of spins as well as majorana fermion operators under symmetry operations are listed in Table \ref{tab:transform} \cite{you2012doping}. Translation symmetry and $C_3(=C_6^2)$ symmetry connect the forms of spins on different sites (on the same sublattice notwithstanding) and for different components, which leads to the  conclusion that spin forms should be ``universal" on all sites of the same sublattice and their components are connected by $\frac{2\pi}{3}$ rotation. If $\sigma,C_6$ symmetries are preserved, we'll be able to connect spins on different sublattices and arrive at similar results (the order of sites in corresponding products on different sublattices might differ notwithstanding). The $h_x$ symmetry leads to some more nontrivial observation, since it can transform a $x$-diectionspin to itself up to a sign. If we are considering a site configuration of the matter Majorana operators to represent $x$-direction spins that's symmetric with respect to the $x$-bond direction, there should be odd number of sites on each side of $x$-bond axis. Write 
\begin{equation}
\sigma^x \sim \eta c_1c_2...c_n c_{n'}...c_{2'} c_{1'} c_{s1} c_{s2}...\end{equation}
where site $i$ and $i'$ are symmetric sites, $c_{sn}$'s are Majorana operators on the symmetric axis and $\eta$ is either $i$ or $1$ to ensure hermiticity. Under $h_x$, it transforms to 
\begin{equation}
\begin{split}
\sigma^x &\rightarrow \eta c_{1'}c_{2'}...c_{n'} c_n...c_2 c_1 c_{s1} c_{s2}...\\
&=\eta (-1)^n c_1c_2...c_n c_{n'}...c_{2'} c_{1'} c_{s1} c_{s2}...
\end{split}
\end{equation}
Since $\sigma^x$ changes sign under $h_x$, $n$ should be odd. So there should at least be one matter Majorana operator on each side both of which combined entertains a symmetric site configuration, which is exactly what the previous form we find in the main text 

\begin{equation}
\tilde\sigma_{i,0}^\mu \sim \frac{i}{2} f \epsilon^{\mu\nu\lambda} c^{\vphantom\dagger}_{i+ s_i \hat{\nu}} c_{i+s_i \hat{\lambda}}
\end{equation} 
looks like. For terms that're asymmetric, $h_x$ requires that their ``images" under this  symmetry also enter the spin form albeit with a minus sign. We use these symmetry constraints to arrive at a general form for the sum of all allowed two-fermion products in the continuum limit below in Eq.~\eqref{eqn:matrix}. We denote $h_\mu$ symmetries as $\pi$ rotations around a $\mu$-direction bond  axis.

From the above analysis of symmetry constraints on the fermion product terms we can write down the most general form of two-fermion products (the symmetry constraints principally manifest in two ways here:  first the two matter Majorana fermions have to live on the same sublattice which is enforced by time-reversal symmetry; and second the ``image" of each present two-fermion product under the corresponding $h_\mu$ operation must also enter the sum albeit with a minus sign or  equivalently, the two-fermion site order exchanged). The most general expression for spin $\sigma_i^{\mu}$ will be a summation of terms like following
\begin{equation}
\label{eqn:general form}
f_{ijk}^\mu\ii(c_{j}c_{k}+c_{h_\mu(k)}c_{h_\mu(j)})
\end{equation}
where $f_{ijk}^\mu$ is some coefficient that depends on the detail of the interaction, sites $j,k$ (whose coordinates are $\bm r_i-\bm l_1,\bm r_i-\bm l_2$ respectively belong to the same sublattice and $h_\mu(j)$'s are reflected images of site $j$'s under $h_\mu$.

Expanding the above expression in terms of the continuum fields, we get
\begin{equation}
\label{eqn:primary}
\begin{split}
&4\psi_a\psi_a^\ast \sin [\bm q_0\cdot \bm \delta]+[\ii\psi_a(\bm \nabla \psi_a \cdot \bm \delta) e^{-\ii\bm q_0\cdot(\bm l_1+\bm l_2)}e^{2\ii\bm q_0\cdot \bm r}\\
&-\ii\psi_a(\bm \nabla \psi_a \cdot h(\bm \delta)) e^{-\ii\bm q_0\cdot [h(\bm l_1)+h(\bm l_2)]}e^{2\ii\bm q_0\cdot \bm r}+\text{h.c.}]
\end{split}
\end{equation}
where $\bm \delta=\bm l_1-\bm l_2$ ($\bm q_0\cdot\bm\delta\equiv-\bm q_0\cdot h_\mu(\bm\delta)\!\mod2\pi$) is the displacement between the sites of the two matter femions, $\bm r$ is the center of the hexagon pertaining to site $i$ (note that the subtlety arising from the difference between the coordinates of the continuum fields and the actual fermion sites is resolved by virtue of $\bm q_0\cdot \hat z=0$). To further simplify the above expression, we have to distinguish between the scenario where the sublattice of the matter fermion is the same/different as that of the spin they constitute (say we consider using $A/B$ sublattice matter fermions to represent an $A$ sublattice spin). We are doing it because $\bm q_0\cdot \bm l\equiv -\bm q_0\cdot h_\mu(\bm l)\!\mod {2\pi}$ ($\bm l \equiv \bm l_1+\bm l_2$) when $\bm l$ is a bravais lattice vector which is true only if the matter fermion sublattice is the same as that of the spin. When in this case, after decomposing $\bm \delta$ into components that're parallel/perpendicular to the $\mu$-bond direction by virtue of the above identity, the general expression is simplified to be
\begin{equation}
\label{eqn:same}
\begin{split}
4&\psi_A\psi_A^\ast\sin [\bm q_0\cdot\bm \delta]+[\ii2e^{2\ii\bm q_0\cdot \bm r}\psi_A\bm \nabla \psi_A\cdot\\
&(\bm\delta _{\perp}\cos [\bm q_0\cdot \bm l]-\bm\delta_{\parallel}\ii\sin [\bm q_0\cdot \bm l])+\text{h.c.}]
\end{split}
\end{equation}
On the other hand, if the matter fermion sublattice is different from that of the spin, we have to modify the displacement $\bm  l_i$ to make it a bravais lattice vector. In order to have a uniform formality, we choose to add a vector $\hat{\bm\mu}$ that corresponds to the $\mu$-bond (from sublattice $A$ to $B$), since $h_\mu(\bm l_i+\bm{\hat \mu})=h_\mu(\bm l_i)+\bm{\hat \mu}$. Rewriting Eq.~\eqref{eqn:primary} with the modification, and decomposing $\bm \delta$ as before, the expression reads ($\bm l=\bm l_1+\bm l_2$)
\begin{equation}
\label{eqn:different}
\begin{split}
&4\psi_B\psi_B^\ast\sin [\bm q_0\cdot\bm \delta]+[\ii e^{2\ii\bm q_0\cdot (\bm {\hat \mu}+\bm r)}2\psi_B\bm \nabla \psi_B \cdot\\
&(\bm\delta _{\perp}\cos [\bm q_0\cdot (\bm l+2\bm{\hat\mu})]-\bm\delta_{\parallel}\ii\sin [\bm q_0\cdot (\bm l+2\bm{\hat\mu})])+\text{h.c.}]\end{split}
\end{equation}
We further have $\bm q_0\cdot\bm {\hat\mu}=\pm\frac{2\pi}{3},0 (\mu=x,y,z)$.  The spin operator forms in Eq.~\eqref{eqn:general form} for three different directions only differ in that the corresponding sites $j,k$ in them are related through $C_3$ rotation, namely all displacements ($\bm \delta,\hat {\bm\mu},\bm l$) in the above simplified expressions are $C_3$ rotation connected. Taking into account that $\bm q_0\cdot \bm l\equiv \bm q_0\cdot C_3(\bm l)\!\mod 2\pi$ if $\bm l$ is a bravais lattice vector, the $\mu$ dependence of corresponding arguments in the trigonometry functions can be dropped. Therefore we have established the connections between spins of different directions on one sublattice site.

As for spins on $B$ sublattice sites, it's obtained from $A$ sublattice spins by the $\sigma$ symmetry transformation. The spin expression in Eq.~\eqref{eqn:general form} transforms into $f_{ijk}^\mu\ii(c_{\sigma(j)}c_{\sigma(k)}+c_{\sigma (h_\mu(k))}c_{\sigma(h_\mu(j))})$. Note that under $\sigma$, spin operators change sign, so the order of the matter fermions is reversed to represent $B$ sublattice spins as $f_{ijk}^\mu\ii(c_{\sigma(k)}c_{\sigma(j)}+c_{\sigma (h_\mu(j))}c_{\sigma(h_\mu(k))})$. So the displacement $\bm \delta_B$ for $B$ sublattice spins equals $-\sigma(\bm \delta_A)$. By virtue of the fact that $\bm q_0\cdot \bm l\equiv \bm q_0\cdot \sigma(\bm l)\!\mod 2\pi$ if $\bm l$ is a bravais lattice vector, we only need to modify $\sin[\bm q_0\cdot\bm \delta]$,$\hat {\bm \mu}$,$\bm \delta _\perp$ to their opposites in Eqs.~\eqref{eqn:same},~\eqref{eqn:different} to represent $B$ sublattice spins. 

Summing over all possible two-fermion product expressions as above, we can write down the expression for $\sigma_i^\mu$. The slowly-varying component and $2\bm q_0$ wavevector varying components are
\begin{eqnarray}
\label{eqn:matrix}
\hat M_a^{\mu}&&=\psi^\dagger m_a\psi\nonumber\\
&&=\psi ^\dagger \left(\begin{array}{cc} \delta_{Aa} m_1+\delta_{Ba} m_2& 0
\\0&-\delta_{Ba}m_1-\delta_{Aa}m_2\end{array}\right)\psi \nonumber\\
\hat N_a^{\mu}&&=\psi^\intercal \bm n_a^{\mu}\cdot\bm \nabla\psi\nonumber\\
=\psi ^\intercal &&\left(\begin{array}{cc} \delta_{Aa}\bm n_1^\mu+\delta_{Ba}(\bm  n_2^{\mu})^{\ast} &0\\
0& -\delta_{Ba}(\bm n_1^{\mu})^{\ast}-\delta_{Aa} \bm n_2 ^\mu \end{array}\right)\cdot \bm \nabla \psi \nonumber\\
\bm n_1^\mu&&=\bm \Delta _{1\perp}^\mu+i\bm \Delta_{1\parallel}^\mu \quad
\bm n_2^\mu=e^{{2i\bm q_0\cdot \bm {\hat \mu}}}(\bm \Delta _{2\perp}^\mu+i\bm \Delta_{2\parallel}^\mu),
\end{eqnarray}
respectively, where $\delta _{ab}$ is kronecker delta function, $m_n\in \mathbb R$ and $\bm \Delta_{n\perp/\parallel}^{\mu}$'s are ``effective" displacement vectors that're perpendicular($\perp$)/parallel($\parallel$) to the $\mu$ bond direction ($\bm \Delta^\mu$'s are related by $C_3$ rotation), which can't be fixed solely on symmetry grounds. 

The symmetry transformation on the continuum fields are as listed in Table \ref{tab:spinor_transform2}. 
 And one can also check that the above minimal form indeed transforms as spin does under the symmetries listed.
\begin{table}
\caption{The transformation of continuum fields under symmetry operations. The transformation of $\psi^\ast$'s will be the conjugate.}
\begin{center}
  \begin{tabular}{@{} cccccc @{}}
    \hline
     operator&& $C_6$ & $\sigma$ & $\mathcal {T}$ & $h_x$ \\ 
    \hline
    $\psi_A$ &$\rightarrow$&  $e^{\frac{-\ii2\pi}{3}}\psi^\ast_B$ & $\psi_B$ & $\psi^\ast_A$ & $-e^{\frac{\ii2\pi}{3}}\psi^\ast_A$ \\ 
    $\psi_B$ &$\rightarrow$&  $-e^{\frac{\ii2\pi}{3}}\psi^\ast_A$ & $-\psi_A$ & $-\psi^\ast_B$ & $-e^{\frac{-\ii2\pi}{3}}\psi^\ast_B$ \\ 
    \hline
  \end{tabular}
\end{center}
\label{tab:spinor_transform2}
\end{table}%

The spin dynamical response in Euclidean space can also be acquired analytically at this isotropic point, we provide for example the low-frequency spin correlations  $S_{ab}^{\mu\mu}(\bm q,i\omega_n)$ for $\bm q\sim 0 $ below
 \begin{equation}
 \begin{split}
S_{ab}^{\mu\mu}(\bm q,\ii\omega_n)\sim \left(\frac{6 \omega_n ^2+3q^2v^2}{\sqrt{q^2 v^2+\omega_n ^2}}\right)(m_1+m_2)^2\nonumber\\ +\left(\frac{6q^2v^2}{\sqrt{q^2 v^2+\omega_n ^2}}\right)\left[\delta_{ab'} (m_1^2+m_2^2)+\delta_{ab}2m_1m_2\right],\nonumber
\end{split}
 \end{equation}
 where $b'$ denotes the complementary sublattice of $b$. For the special simple spin forms we find in Eq.~\eqref{eqn:simple}, the parameters in the spin forms in Eq.~\eqref{eqn:matrix} are
 \begin{eqnarray}
 m_1&&=0\quad m_2=2\sqrt{3}\nonumber\\
 \bm n_1^\mu&&=\left (0,0\right)\quad
 \bm n_2^\mu=-2\sqrt{3}e^{\ii2\bm q_0\cdot \hat {\bm \mu}}\left(\sin [2\bm q_0\cdot\hat{\bm\mu}],\cos[2\bm q_0\cdot\hat{\bm\mu}]\right).\nonumber
 \end{eqnarray}
 We further have for spin correlations (take $v=1$)
 \begin{eqnarray}
 &&\sum_\mu S_{AA/BB}^{\mu\mu}(\bm q\sim 0,\ii\omega_n)\sim \frac{q^2+2\omega_n^2}{\sqrt{\omega_n^2+q^2}}\nonumber\\
 && \sum_\mu S_{AB/BA}^{\mu\mu}(\bm q\sim 0,\ii\omega_n)\sim \frac{3q^2+2\omega_n^2}{\sqrt{\omega_n^2+q^2}}
  \nonumber\\
  && \sum_\mu S_{AA/BB}^{\mu\mu}(\bm q=2\bm q_0+\bm k,\ii\omega_n)\sim 2\frac{k^4+8\omega_n^2k^2+8\omega_n^4}{\sqrt{\omega_n^2+k^2}}
  \nonumber\\
  &&  \sum_\mu S_{AB/BA}^{\mu\mu}(\bm q=2\bm q_0+\bm k,\ii\omega_n)\sim -\frac{-k^4+8\omega_n^2k^2+8\omega_n^4}{\sqrt{\omega_n^2+k^2}}\nonumber
    \end{eqnarray}
    and the asymptotic behavior is conspicuously exhibited.
    
The spectral function for this special form which we have also numerically calculated henceforth reads
\begin{eqnarray}
&&A_S(\bm q\approx 0, \omega)\sim \text{sgn}(\omega)\theta(|\omega|- |\bm q|) \frac{4\omega^2-4q^2}{\sqrt{\omega^2-q^2}},\nonumber\\
&&A_S(\bm q=2\bm q_0+\bm k, \omega)\sim \text{sgn}(\omega)\theta(|\omega|-|\bm k|) \frac{3k^4-8\omega^2k^2+8\omega^4}{\sqrt{\omega^2-k^2}}.\nonumber
\end{eqnarray}
where sgn$(x)$, $\theta(x)$ denote the sign and Heaviside step function, respectively.

\section{Appendix B: Vanishing of low order contributions to $\tilde\sigma_{i,0}^\mu$, and structure of the flux analysis via unitary transformation}

Let us consider how low-energy expansions of the spin operator arise from specific microscopic perturbations to the Kitaev model by first tracing the fate of the fluxes. 
We have illustrated in the main text that the transformed spin operator has an expression of the form
\begin{eqnarray}
\tilde\sigma _i^{\mu}=\underbrace{\ii Zc_ic_i^\mu+\cdots}_{\tilde\sigma^\mu_{i,2k>0}}+\underbrace{ f^\mu_{ijk} \ii c_{j}c_{k}+\cdots}_{ \tilde\sigma^\mu_{i,0}}\; ,
\end{eqnarray}
where $Z=1$ for the ideal Kitaev model but is reduced by perturbation. 
To obtain the form given in the second bracketed terms of the right hand side in Eq. \eqref{eqn:general}, we first set aside the complicated spinon configuration induced by matter fermion proliferation during the perturbation process, and fix our attention on the flux sector, since these are static in the Kitaev limit. The annihilation of all the fluxes is a \emph{ necessary but not sufficient} condition for the existence of contributions to $\tilde\sigma_{i,0}^\mu$.

\begin{figure}[htbp]
\begin{center}
\includegraphics[width=0.35\textwidth]{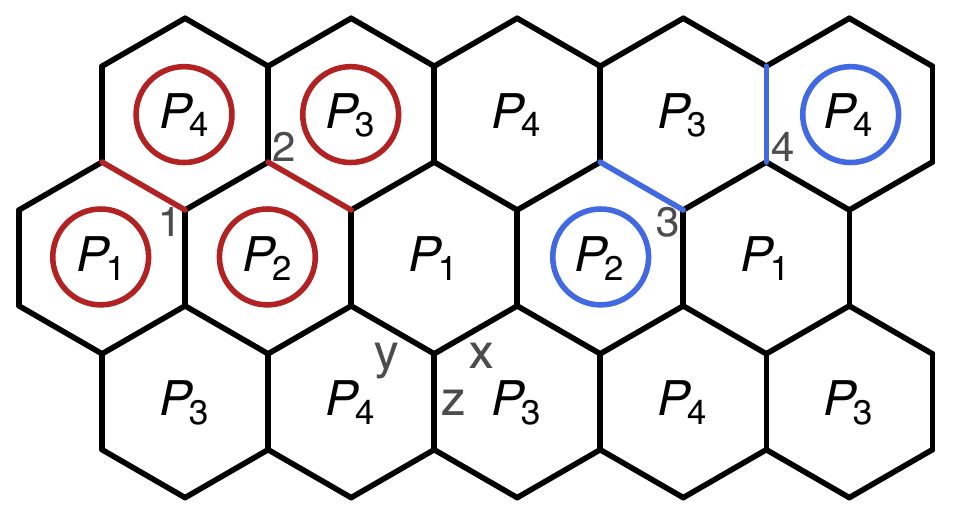}
\caption{The action of $\sigma_1^y\sigma_2^y$ on the ground state creates four fluxes, shown as red circles, because the corresponding red bond values are flipped. The plaquettes are partitioned into four types labeled by $P_1$-$P_4$. The Heisenberg interaction simultaneously changes the flux number by one in all four types of plaquettes. The action of the cross term such as $\sigma_3^y\sigma_4^z$ flips the blue bond values and creates two fluxes, drawn as blue circles in the plaquettes linked by bond 3-4. If we apply these two terms consecutively on the same x-bond, the remaining fluxes are just be the ones which would be created by corresponding x component of spin.  
}
\label{fig:heisenberg_term}
\end{center}
\end{figure}

We begin with the isotropic Heisenberg interaction, $V_H=J_H\sum_{\langle ij\rangle} {\bm {\sigma_i }}\cdot {\bm{\sigma_j}}$, which, together with the Kitaev term, defines the Kitaev-Heisenberg model introduced early on for the honeycomb family of iridium oxides \cite {Chaloupka2010}.  The term $\sigma ^\mu_i\sigma ^\mu _j$ with $\mu$ different from the $\langle ij\rangle$ bond direction modifies the flux in the four plaquettes sharing sites $i,j$ (as marked by red circles in Fig.~\ref{fig:heisenberg_term}), so that it contributes to $U_4$ in Eq. (4) of the main text to leading order.  One may imagine that higher order terms can induce $U_2$ terms and hence the desired flux-free contribution to $\tilde\sigma_i^\mu$.  However, this is not the case, and flux-free contributions are absent at all orders for the pure $V_H$ perturbation.  To see this,  we introduce a labeling of plaquettes into four types $P_1\text -P_4$ as shown in Fig.~\ref{fig:heisenberg_term}.   The Heisenberg interaction $V_H$ simultaneously changes the parity of the flux in all four sets of plaquettes, while each spin operator only changes the flux parity in two of the four sets.  In other words, the Heisenberg interaction conserves the product of any two flux parities, unlike the spin operator.  This implies that states created by the spin operator cannot mix with the low energy sector.  The pair flux parity conservation is in fact equivalent to a discrete global ``dihedral'' symmetry:  the quantities $\prod_{i} \sigma^\mu_i (\mu=x,y,z)$ commute with $H_0$ and $V_H$.  For example, i.e., the total flux parity in $P_1,P_2$-type plaquettes is $\prod_{p=P_1,P_2\text {-}type} W_p = \prod_{i} \sigma^z_i$.   The spin operator does not commute with all three global operators and therefore cannot transform into pure matter fermion excitations that conserve these quantities. We note that this observation is also made in Ref.\onlinecite{mandal2011sf}.

\begin{figure}[tbp]
\begin{center}
\includegraphics[width=0.49\textwidth]{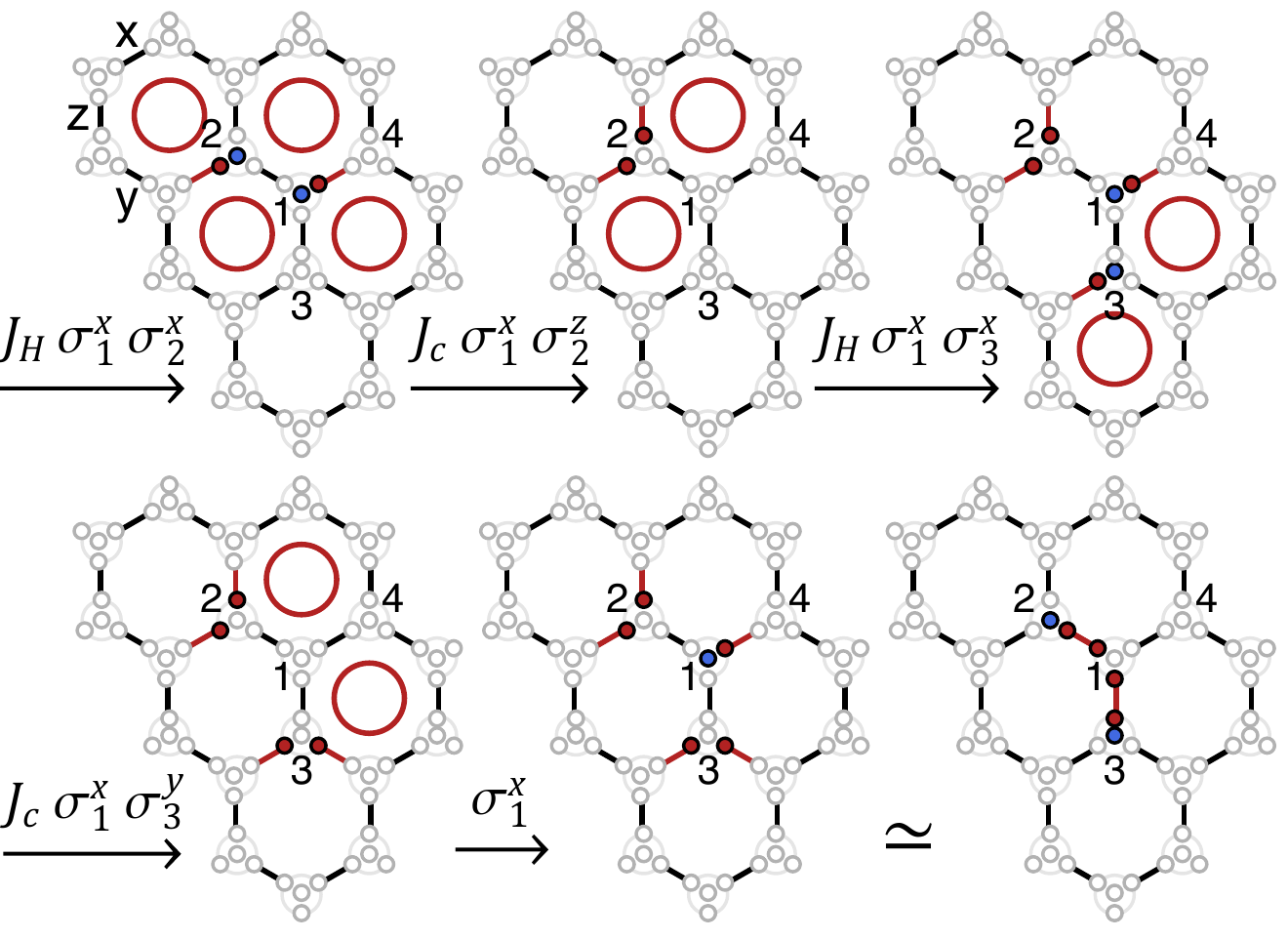}
\caption{The $4^{\rm th}$ order perturbation to $\sigma_1^x$ that results in the spin form $ic_2c_3$. The matter fermion operators are denoted as blue dots, and the bond Majorana fermion operators are denoted as red dots. The fluxes are represented as circles in hexagons. The last step rearranges the fermion configuration by virtue of the gauge constraint $c_ic_i^xc_i^yc_i^z=1$, which results in the desired form of the product of fermion operators with gauge strings connecting them.}
\label{fig:4th order}
\end{center}
\end{figure}
However, the generic spin model\cite{Rau2014} for the Kitaev materials does not possess dihedral symmetry, which is for example broken by the bond-dependent symmetric off-diagonal exchange (cross term) $V_c=J_c\sum_{\mu(\nu\gamma)}\sum_{\langle ij\rangle ^\mu}(\sigma^\nu_i\sigma^\gamma_j+\sigma^\gamma_i\sigma^\nu_j)$ ($\nu,\gamma$ are the remaining directions). The cross term acting on a bond creates two fluxes in the plaquettes that are {\em linked} by that bond (as marked by blue circles in Fig.~\ref{fig:heisenberg_term}).  On its own, by arguments similar to above (but with a different partition of the plaquettes that is not shown here), $V_c$ cannot annihilate the fluxes created by a single spin.  We note that this result shows, by counterexample, the incorrectness of the claim of Ref.\onlinecite{mandal2011sf} that Eq.(50) of that paper is a necessary and sufficient condition for the spin correlators to stay short-ranged.  However, together with the Heisenberg exchange, it is possible to induce an appropriate set of virtual fluxes in the ground state that allows a flux-free contribution to $\tilde\sigma_i^\mu$: applying one factor of the Heisenberg interaction on the same bond as one from the cross term leaves two fluxes on the plaquettes adjacent to the bond, which is the same configuration as that of a bare spin operator.  So we seem to arrive by $O(V_H V_c)$  at a combination of perturbations that can annihilate the spin-created fluxes. However, it turns out that this does not contribute to $\tilde\sigma_i^\mu$.  To see this, we consider the product of spin operators appearing in the corresponding perturbation expansion for $\tilde\sigma_1^x \sim  ({\sigma^y_1\sigma^z_2})({\sigma^y_1\sigma^y_2})\sigma^x_1 + \cdots$, which, by simple algebra gives $\tilde\sigma_1^x \sim c_2c_1 + \cdots$ (here $1,2$ are neighboring sites that share a x-bond, as illustrated in Fig.~\ref{fig:heisenberg_term}).  This term is {\sl anti-Hermitian} and so must vanish after including the contribution of the Hermitian conjugate counterpart, which one can verify arises in the unitary transformation formula.  Thus we find that the second order contributions to $\tilde\sigma_{i,0}^\mu$ vanish.  The first non-vanishing contribution occurs, as discussed in the main text, at fourth order.  (See fig \ref{fig:4th order})

\bibliographystyle{apsrev4-1}
%

\end{document}